# THE ROLE OF COMMUNICATION IN EFFECTIVE BUSINESS MANAGEMENT

Dariusz BARAN[1*], Ernest GÓRKA[2], Michał ĆWIĄKAŁA[3], Gabriela WOJAK[4], Mateusz GRZELAK[5], Katarzyna OLSZYŃSKA[6], Piotr MRZYGŁÓD[7], Maciej FRASUNKIEWICZ[8], Piotr RĘCZAJSKI[9], Maciej ŚLUSARCZYK[10], Jan PIWNIK[11]

[1] Pomorska Szkoła Wyższa w Starogardzie Gdańskim, Instytut Zarządzania, Ekonomii i Logistyki; dariusz.baran@twojestudia.pl, ORCID: 0009-0006-8697-5459
[2] Wyższa Szkoła Kształcenia Zawodowego; ernest.gorka@wskz.pl, ORCID: 0009-0006-3293-5670
[3] Wyższa Szkoła Kształcenia Zawodowego; michal.cwiakala@wskz.pl, ORCID: 0000-0001-9706-864X
[4] I'M Brand Institute sp. z o.o.; g.wojak@imbrandinstitute.com, ORCID: 0009-0003-2958-365X
[5] I'M Brand Institute sp. z o.o.; m.grzelak@imbrandinstitute.com, ORCID: 0009-0000-1395-3087
[6] Polsko Japońska Akademia Technik Komputerowych; kontakt@olszynska.com, ORCID: 0009-0003-4309-6233
[7] Piotr Mrzygłód Sprzedaż-Marketing-Consulting; piotr@marketing-sprzedaz.pl, ORCID: 0009-0006-5269-0359
[8] F3-TFS sp. z o.o.; m.frasunkiewicz@imbrandinstitute.com, ORCID: 0009-0006-6079-4924
[9] MAMASTUDIO Pawlik, Ręczajski, sp. j.; piotr@mamastudio.pl, ORCID: 0009-0000-4745-5940
[10] Pomorska Szkoła Wyższa w Starogardzie Gdańskim, Instytut Zarządzania, Ekonomii i Logistyki; maciej.slusarczyk@twojestudia.pl, ORCID: 0000-0001-6612-8179
[11] WSB Merito University in Gdańsk, Faculty of Computer Science and New Technologies; jpiwnik@wsb.gda.pl, ORCID: 0000-0001-9436-7142
* Correspondence author

**Purpose:** This study aims to investigate the role of communication in effective business management by comparing internal communication systems in two companies, focusing on managerial practices, employee perceptions, and communication tools.
**Design/methodology/approach**: A quantitative survey of 220 employees assessing 15 aspects of internal communication. Statistical methods identified key differences and patterns.
**Findings:** Company X outperforms Company Y in all communication areas, showing higher clarity, better feedback culture, and stronger use of technology.
**Research limitations/implications**: The study is limited to two companies in one sector and country. Results are based on self-reported data.
**Practical implications:** Investing in feedback systems, digital tools, and participatory models improve communication effectiveness.
**Social implications:** Strong internal communication supports trust, inclusion, and job satisfaction.
**Originality/value:** The study offers rare, data-driven comparison of communication practices in similar companies within one industry.
**Category of the paper:** research paper.

## 1. Introduction

In the era of dynamic technological transformation and rising expectations toward organizational agility, internal communication has become a strategic function within companies. Far beyond the mere exchange of information, it plays a critical role in shaping corporate culture, fostering employee engagement, and enabling effective management. Despite its acknowledged importance, there remains a practical gap in understanding how internal communication systems operate in comparable organizational contexts, particularly within the same industry.

To address this gap, this article presents a comparative study of internal communication practices in two Polish car rental companies - referred to as Company X and Company Y. These are fictitious names used for the purpose of this research to protect confidentiality and generalize findings. However, the organizational data provided reflects realistic operational profiles of medium-sized enterprises in the car rental industry.

Company X is headquartered in Warsaw and employs approximately 150 people. With branches in cities such as Kraków, Wrocław, Poznań, Gdańsk, and Katowice, the company offers a wide range of vehicles, including passenger cars, SUVs, and luxury models. It emphasizes high service standards, technological innovation, and sustainability, including an expanding fleet of electric vehicles. The company maintains a strong focus on customer experience, operating a loyalty program and online booking via both a website and mobile application. It is known for its advanced organizational structure, with dedicated departments for innovation, finance, customer service, IT, and legal compliance.

Company Y is located near Warsaw (Łomianki) and employs approximately 170 people, with regional branches in cities such as Gdańsk, Kraków, Bydgoszcz, and Płock. While its fleet and services are comparable to Company X, it places a stronger emphasis on operational efficiency, vehicle maintenance, and rapid customer support. The company operates its own vehicle maintenance center and car wash, ensuring high fleet readiness. It also runs a loyalty program and focuses on individual client support, offering 24/7 roadside assistance and flexible leasing options.

The aim of this study was to assess the perceived effectiveness of managerial communication tools, identify communication barriers, and evaluate employee involvement in internal information processes. Based on a quantitative diagnostic survey conducted among 220 employees using a structured questionnaire, the study provides statistically significant insights into how different approaches to communication influence operational efficiency and employee satisfaction.

The originality of this study lies in its direct, data-driven comparison of two similar enterprises operating under shared market conditions. This article extends the current state of knowledge by applying a comparative lens to two distinct organizational models within the same industry.

The findings confirm that Company X, which utilizes technologically advanced tools and promotes participatory communication models, achieves higher employee ratings across all key dimensions, including feedback culture, message clarity, and interdepartmental coordination. In contrast, Company Y shows greater inconsistency and weaker communication practices, particularly in terms of feedback regularity and transparency. These results reinforce the theoretical assumption that effective, bidirectional communication strengthens employee engagement, strategic alignment, and overall organizational performance.

## 2. Internal communication: concepts, functions, and empirical insights

Communication is considered one of the fundamental components of effective business management. As John Adair (2010) notes, the term "communication" originates from the Latin word *communicatio*, which means "sharing" or "making common". In the business context, communication goes far beyond the simple exchange of information—it becomes a key tool for building trust, engaging employees, and navigating organizational change (Griffin, 2007). It allows companies to create a shared vision and ensure all employees understand the goals and mission of the enterprise.

According to Mała and Hrabelska (2013), communication is a complex system that encompasses cognitive, affective, formal, and evaluative components. It plays a critical role in coordinating actions, motivating employees, and fostering collaboration across teams. Effective communication systems within organizations ensure the smooth flow of information both internally and externally, shaping relations with employees, customers, and business partners alike.

Griffin (2007) defines effective communication as the process of sending a message in such a way that the meaning received is as close as possible to the meaning intended. Stoner, Freeman, and Gilbert (2011) extend this understanding by stating that communication involves people striving to share meanings through symbolic messages. This implies the need for mutual understanding of symbols and the development of a shared context between sender and receiver.

Within the managerial context, communication supports four key functions: planning, organizing, motivating, and controlling (Stankiewicz, 2006). Planning involves the communication of goals, forecasts, and operational methods; organizing includes forming cohesive teams and coordinating efforts; motivating relies on emotional intelligence, understanding employee needs, and providing feedback; and controlling involves clear and objective delivery of expectations and standards.

Mała and Hrabelska (2013) identify five primary functions of communication: informational, motivational, regulatory, emotional, and social. These functions enable managers to provide direction, clarify roles, encourage engagement, maintain discipline, and foster a sense of belonging. The motivational role of communication, in particular, facilitates the alignment of individual efforts with organizational goals.

From the perspective of organizational culture, communication is also a reflection of shared values, expectations, and behavioral norms. As highlighted by Stankiewicz (2006), the style of communication within a company has a direct influence on employee relations, customer treatment, and the overall atmosphere in the workplace.

An important distinction within organizational communication concerns its form, as outlined by R.W. Griffin (2007), who identifies several types: verbal, non-verbal, oral, written, direct, and indirect. Verbal communication encompasses both spoken and written language—ranging from face-to-face conversations, business meetings, and phone calls to emails and official reports. It allows for the fast and accurate transmission of large amounts of information, especially in high-pressure or decision-making contexts. However, as Griffin emphasizes, verbal communication may also be a source of misunderstandings when the sender and receiver interpret the message differently or use ambiguous language.

Griffin also underlines the significance of non-verbal communication, which conveys emotions, attitudes, and intentions through body language, gestures, facial expressions, and tone of voice. These signals, though subtle, can powerfully reinforce or contradict verbal messages. Oral communication is valued for its immediacy, flexibility, and ability to build social bonds within teams, while written communication offers precision and durability but lacks real-time feedback. Moreover, Griffin distinguishes between direct communication, which relies on personal contact, and indirect communication, which uses intermediaries or technologies like emails, phone calls, or mass media. Each of these forms has its place depending on the organizational context, urgency, and interpersonal dynamics.

Intra-organizational communication can be categorized based on its directional flow. According to Gołuchowski and Filipczyk (2020), vertical communication occurs when messages are transmitted through hierarchical levels—from top to bottom and vice versa. Downward communication includes transmitting company goals, procedures, or decisions from management to employees, while upward communication enables feedback, suggestions, and reports to flow back to leadership. For communication to be effective in a modern organizational context, it must be bidirectional, allowing for feedback and adjustment of the message if necessary. As Łukasiewicz and Pietrzak (2024) argue, this type of reciprocal exchange enhances message accuracy and clarity and helps build procedural security within organizations.

Effective communication, both internally and externally, is a key success factor for businesses - especially in competitive and dynamic environments. As Ćwiklińska (2005) emphasizes, communication must deliver information that is mutually beneficial and clearly understood by all parties involved. One of the fundamental goals of internal communication is to foster strong employer–employee relations. As Maruszak (2014) points out, internal communication should reach all levels of personnel, including trade unions, shareholders, and even employees' families. Its role is to ensure the transparent dissemination of information, which facilitates understanding of the company's values, management strategies, and workplace expectations.

According to Kim (2010), communication that supports effectiveness and positive relationships is marked by trust, job satisfaction, a sense of security, optimism about the future, and belief in collective success. Bidirectional communication is central to achieving these elements.

However, internal communication is not without challenges. Majka-Rostek (2010) warns that horizontal communication may lead to excessive information flow, increasing the risk of "informational noise". In competitive environments, some employees may even be reluctant to assist colleagues, undermining team cohesion.

To sum up, communication within an organization is a multifaceted process that encompasses various forms, directions, and functions. It plays a vital role not only in operational efficiency and employee engagement but also in shaping organizational culture and values. Both vertical and horizontal communication flows, along with verbal and non-verbal elements, contribute to the development of mutual trust, loyalty, and a sense of shared purpose. At the same time, communication challenges such as information overload or hierarchical distortion can hinder cooperation and productivity. Therefore, establishing effective, two-way communication systems remains a priority for modern organizations.

In recent years, numerous studies have emphasized the importance of effective internal communication for organizational performance. For example, research by Psico Smart (2023) indicates that organizations with strong internal communication are 4.5 times more likely to retain top talent, highlighting its strategic relevance. Similarly, the "State of the Sector 2023/24" report by Gallagher presents global trends and key challenges in internal communication practices across industries.

A valuable contribution to the field of internal communication comes from the study by Eskelinen, Rajahonka, Villman, and Santti (2017), who examined two Finnish SMEs participating in a service design training program aimed at improving communication management. Their research applied participative business model techniques to identify and address internal communication challenges between departments such as production, marketing, and delivery. Using the CIMO logic framework (context, intervention, mechanism, output), they demonstrated how involving employees in the communication development process increased motivation, revealed tacit knowledge, and improved organizational alignment. Although their findings showed significant improvements in information flow and strategic communication within each organization, the study did not perform a comparative cross-company analysis. Laframboise & Nelson (2016) Therefore, our study complements and extends their work by investigating internal communication systems in two distinct enterprises - X and Y -allowing for a direct comparison of solutions, challenges, and outcomes in different organizational settings.

Another significant study that reinforces the value of internal communication in organizational settings is that of Santos, Santos, Sousa, and Oliveira (2024), who analyzed the relationship between internal communication, employee motivation, and job satisfaction across 426 employees in Portuguese organizations. Using Partial Least Squares Structural Equation Modeling (PLS-SEM), the authors demonstrated that internal communication exerts both a direct and indirect influence on job satisfaction - mediated by employee motivation. Their findings suggest that effective internal communication not only enhances informational transparency but also fosters emotional engagement and personal fulfillment among employees. This research provides a robust empirical model to support the view that motivation acts as a bridge linking communication quality to overall job satisfaction. Unlike our study, which compares communication systems in two distinct companies (X and Y), their approach focuses on broader organizational patterns across sectors. Nevertheless, their emphasis on two-way communication, managerial openness, and clarity of feedback mechanisms provides a strong theoretical underpinning for our comparative analysis of enterprise-level communication effectiveness.

## 3. Research methodology and case description

Today's organisations, regardless of industry, face the challenge of ensuring efficient, effective and two-way internal communication. It is communication between employees and their superiors that is one of the key elements influencing the quality of cooperation, level of commitment and overall work efficiency. This is of particular importance in companies operating in the service sector, where interpersonal relations and the rapid flow of information are essential for the provision of high-quality services.

This study is devoted to analysing the internal communication system in two companies in the car rental industry - companies X and Y. The subject of the study was the communication relationship between employees and management, focusing on the evaluation of the effectiveness of the solutions applied in these companies. In the context of increasing customer expectations and dynamic technological changes, effective communication is becoming not only a management tool, but also a factor building a company's competitive advantage.

Companies X and Y are two fast-growing companies operating in the car rental sector, consolidating their position in the domestic market between 2020 and 2022 and, in the case of X, also abroad. Both companies offer modern, flexible vehicle rental services, responding to the growing demands of individual and business customers.

Company X was founded in 2008 by a group of entrepreneurs coming from the engineering sector. Since its inception, it has focused on providing comprehensive car rental services to individual and corporate customers. The company is headquartered in Warsaw, while its operational network includes numerous branches in major Polish cities. In 2022, the company began expanding into foreign markets, thus becoming an international player.

Company X's range includes a diverse fleet of vehicles - from economy models to luxury SUVs and limousines, both cars and vans. Vehicles are updated annually to maintain a high technical and aesthetic standard. The company places particular emphasis on the development of additional services (insurance, GPS, child seats, roof racks, snow chains), as well as the flexibility of the offer (short- and long-term rental, cross-border rental).

Company Y was founded in 2010 as an initiative of friends with experience in the automotive sector. The company's head office is located near Warsaw. The company operates throughout Poland through branches in other cities.

Like X, Company Y offers a wide range of vehicles for hire, including modern passenger cars of various classes and luxury vehicles. The fleet is maintained in good technical and aesthetic condition, and the company has its own car wash and vehicle cosmetics department to ensure the high visual quality of the cars on offer.

Y's main strengths revolve around flexibility and a personalised customer approach. The company offers an extensive package of additional services (insurance, GPS, child seats, local transfers, travel equipment), as well as 24/7 technical assistance including towing, fuel supply and replacement cars. An important element of the strategy is the company's proprietary loyalty programme - a customer who hires a car five times during the year receives the sixth rental free of charge (up to five working days), regardless of the class of vehicle.

Both companies have extensive organisational structures, responding to the complexity of the vehicle rental processes. At Company X, the structure is made up of management teams responsible for the various business areas: operations, finance, marketing, technology and innovation, service, human resources and legal issues. The organisation of work is based on central management using information systems, which supports the operational efficiency and scalability of the business model. The Y structure focuses more on direct operational management. The president oversees key logistics and customer service processes, while individual directors manage the operations, IT, technical service, marketing and finance departments. The company also has reservations, leasing and field offices, which ensures proximity to the customer and the ability to respond quickly to market needs.

The main objective of the study is to analyse and evaluate selected communication tools and mechanisms used by managers of companies X and Y in their contacts with employees, as well as to identify possible barriers and areas for improvement.

The research used a quantitative diagnostic survey method, and the main data collection technique was a questionnaire, which is a tool that allows data to be collected directly from respondents, enabling them to make a free but focused statement about the phenomenon under study. The questionnaire of the survey was constructed following the principles of logical and factual correctness - the questions were formulated in a way that was understandable, unambiguous and relevant to the respondents' level of knowledge and experience.

The questionnaire consisted of three parts:

1. Introductory part - including information about the purpose and nature of the study, together with a request to complete the questionnaire honestly and sincerely.
2. Substantive part - including questions on the assessment of the quality of communication between employees and supervisors in the workplace.
3. Metrics and identification section - including demographic and occupational data such as gender, age, education level and position held.

Given the subjective nature of the answers given by the respondents, their interpretation was carried out with caution and taking into account the possibility of measurement error. The analysis of the results focused on the identification of general trends and differences occurring between the two companies, which made it possible to formulate practical conclusions and recommendations for the improvement of internal communication in the surveyed companies.

## 4. Research results

The data presented in Figures 1 to 15 show the responses of 220 employees of two companies in the car rental industry - company X and company Y - regarding various aspects of internal communication. The questionnaire included 15 questions related to, among other things, clarity of communication, accessibility of supervisors, regularity of meetings, flow of information between departments, access to modern tools and provision of feedback.

The responses show differences in the ratings of individual areas between the two companies - in some cases, one of the companies obtained higher percentages of positive answers, in others the differences were less pronounced. The distribution of responses makes it possible to observe how employees rate the quality of communication at operational, managerial and strategic levels. In many of the questions, respondents also referred to specific communication tools and daily information practice in the organisation.

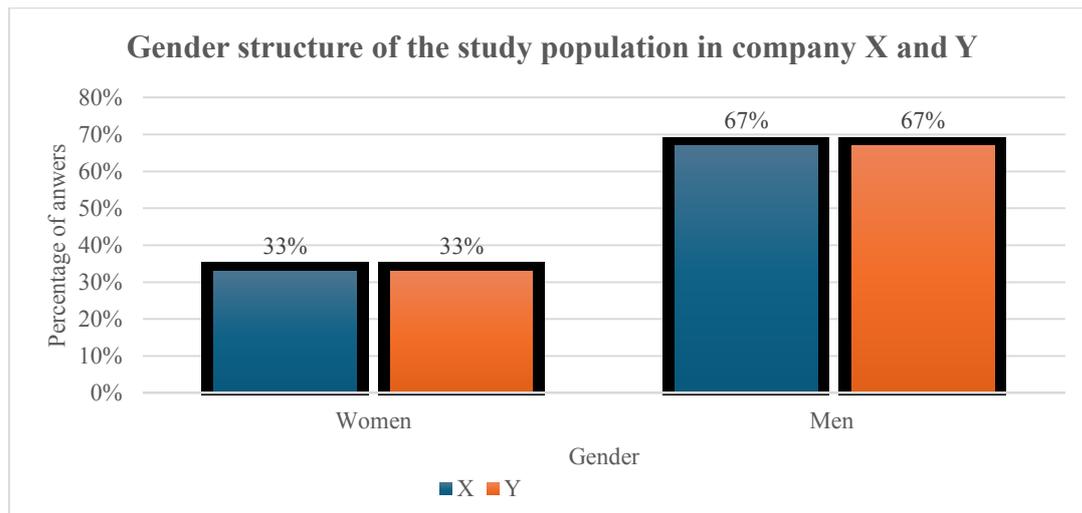

**Figure 1.** Responses of employees of companies that were analysed on the internal communication system to question 1 of the research survey.

In both organizations, men constituted 67% of the study population, while women accounted for 33%, indicating an identical gender structure. This balanced distribution ensures the comparability of gender-based perceptions in the subsequent analysis.

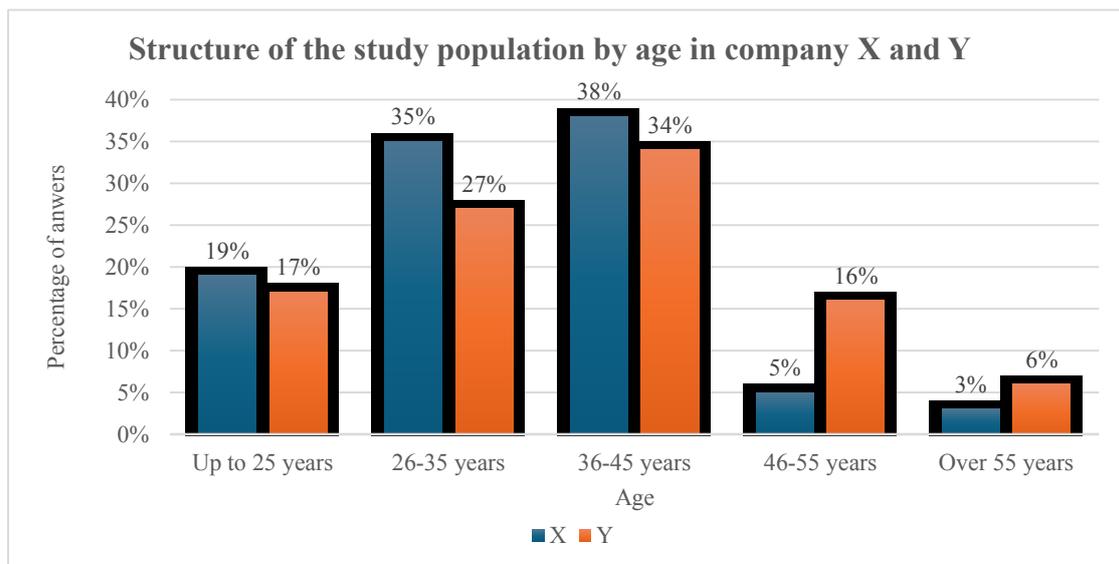

**Figure 2.** Responses of employees of companies that were analysed on the internal communication.

In both companies, the majority of employees are aged 26-45, with Company X having slightly more respondents aged 26-35 (35%) and 36-45 (38%) compared to Company Y (27% and 34%, respectively). Notably, Company Y has a higher share of older employees aged 46 and above, which may influence communication preferences and practices in the organization.

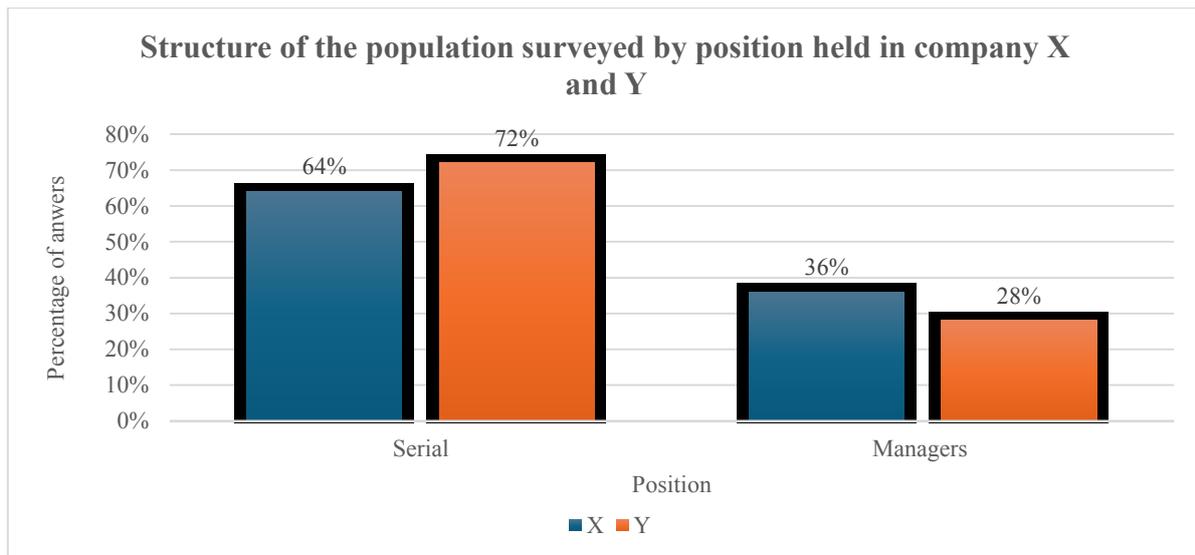

**Figure 3.** Responses of employees of companies that were analysed on the internal communication system to question 3 of the research survey. system to question 2 of the research survey.

In Company Y, 72% of the respondents were non-managerial staff, compared to 64% in Company X. Conversely, Company X had a higher proportion of managers (36%) than Company Y (28%), which may contribute to different perceptions of communication practices across hierarchical levels.

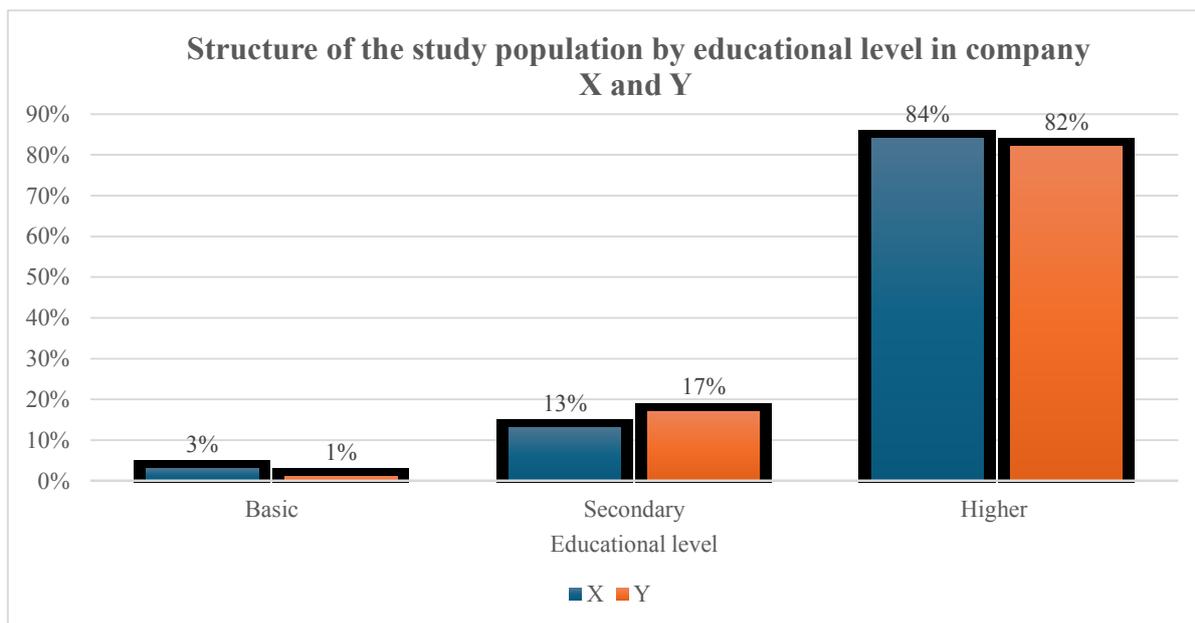

**Figure 4.** Responses of employees of companies that were analysed on the internal communication system to question 4 of the research survey.

In Company X, 3% of respondents had basic education, 13% had secondary education, and 84% had higher education. In Company Y, 1% had basic education, 17% had secondary education, and 82% had higher education. The majority of respondents in both companies held higher education degrees.

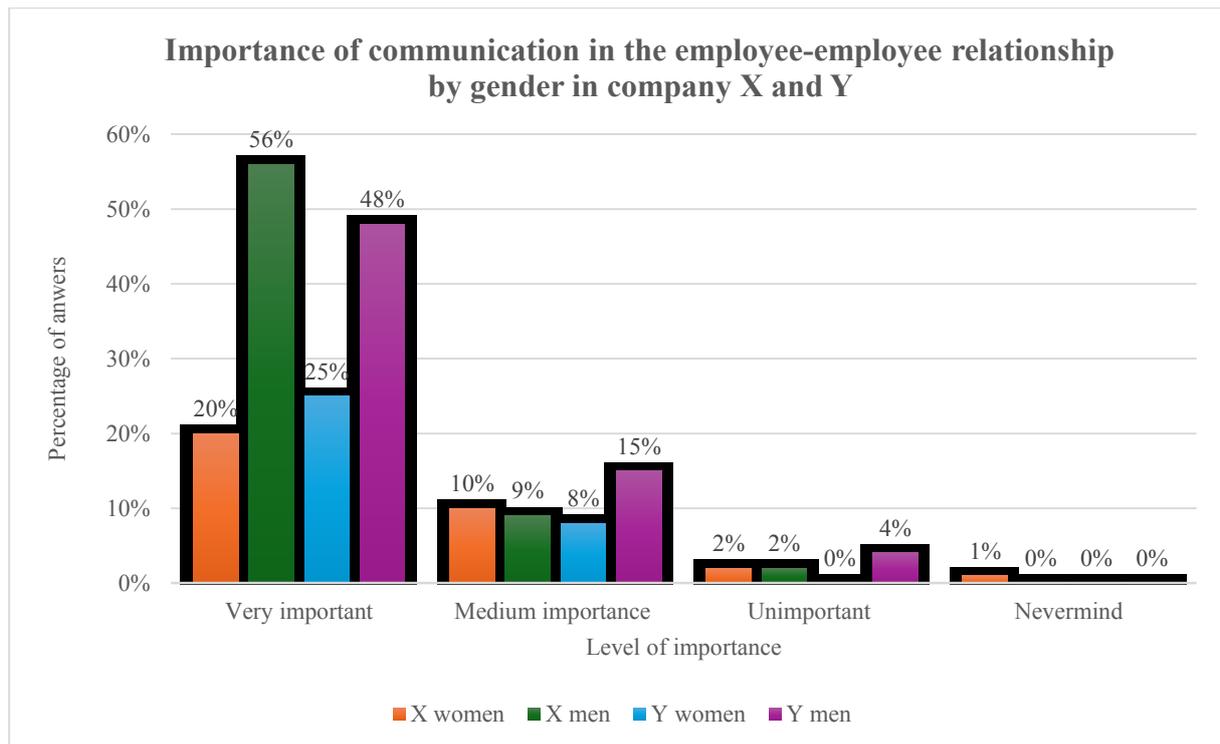

**Figure 5.** Responses of employees of companies that were analysed on the internal communication system to question 5 of the research survey.

Among respondents, 56% of men from Company X and 48% of men from Company Y rated it as "very important," while women from both companies rated it lower, at 20% (X) and 25% (Y). Medium importance was selected by 10% of women in Company X, 9% of men in Company X, 8% of women in Company Y, and 15% of men in Company Y. Only a small share of respondents considered communication unimportant or irrelevant, with values ranging from 0% to 4%.

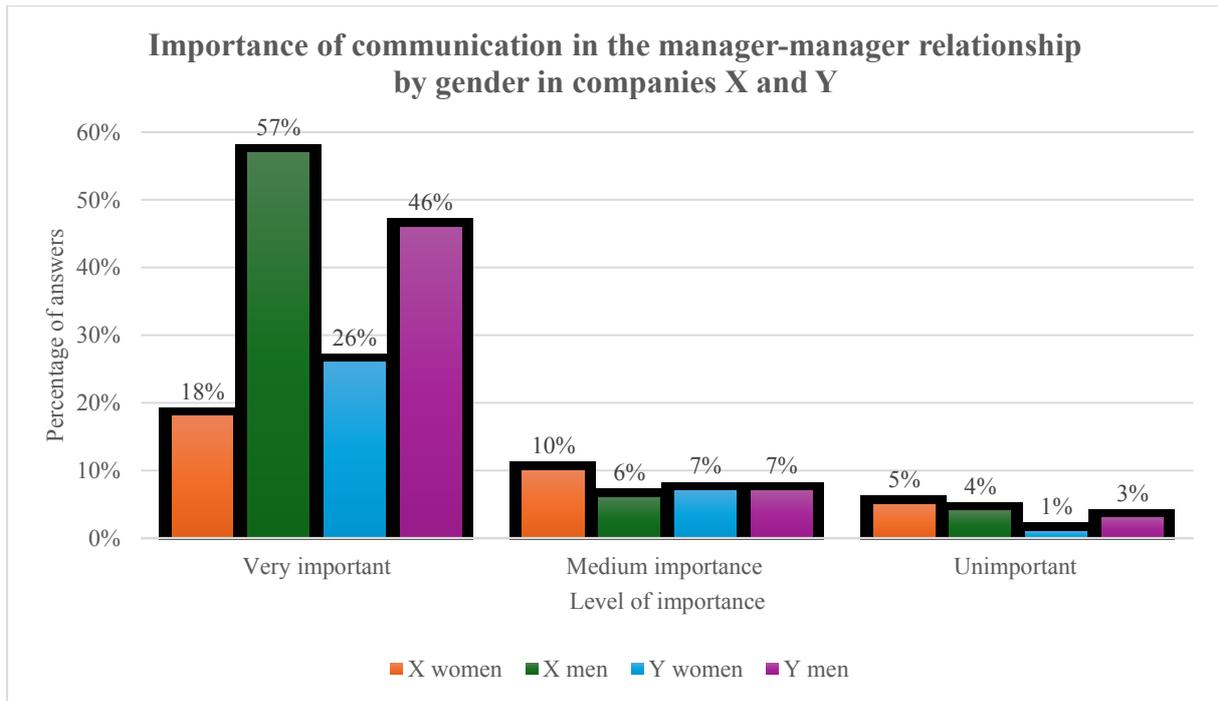

**Figure 6.** Responses of employees of companies that were analysed on the internal communication system to question 6 of the research survey.

Among men, 57% from Company X and 46% from Company Y rated it as "very important," compared to 18% of women from Company X and 26% from Company Y. Medium importance was indicated by 10% of women from Company X, 6% of men from Company X, and 7% of both women and men from Company Y. Only a small proportion considered it unimportant, with values ranging from 1% to 5%.

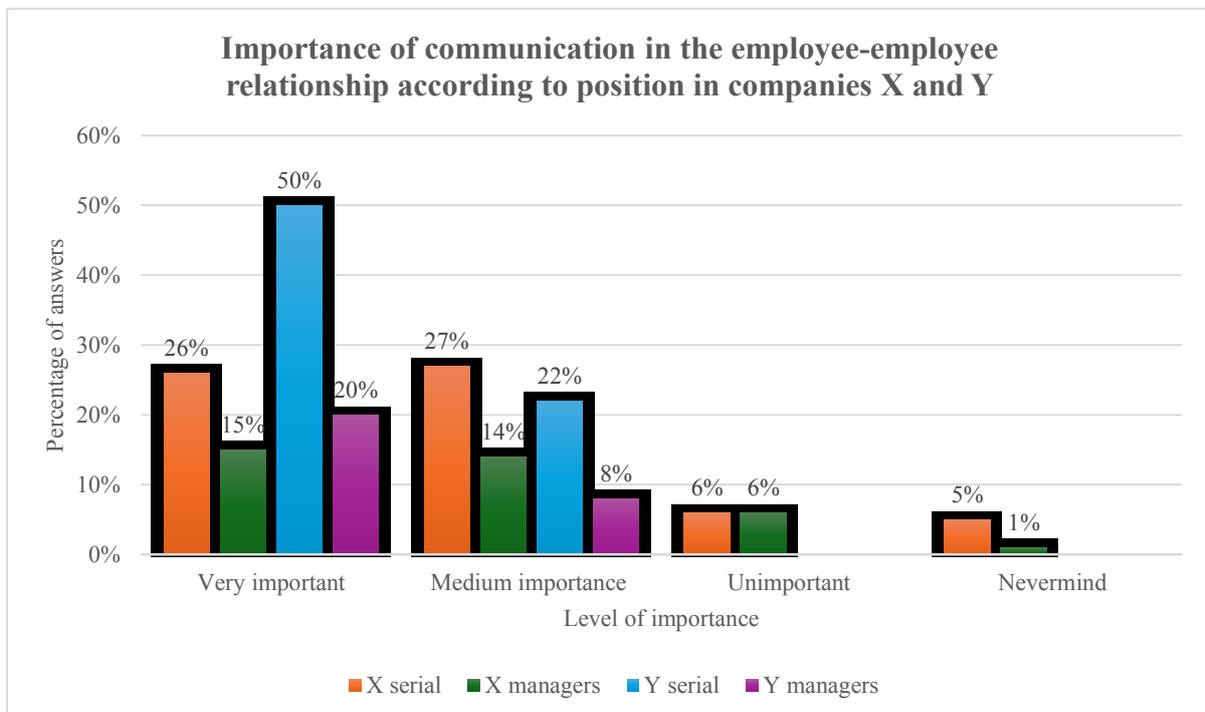

**Figure 7.** Responses of employees of companies that were analysed on the internal communication system to question 7 of the research survey.

Among serial employees, 26% in Company X and 50% in Company Y considered it "very important," while the respective values for managers were 15% in X and 20% in Y. For the "medium importance" category, the results were 27% (X serial), 14% (X managers), 22% (Y serial), and 8% (Y managers). The remaining responses, including "unimportant" and "nevermind," accounted for under 6% in each subgroup.

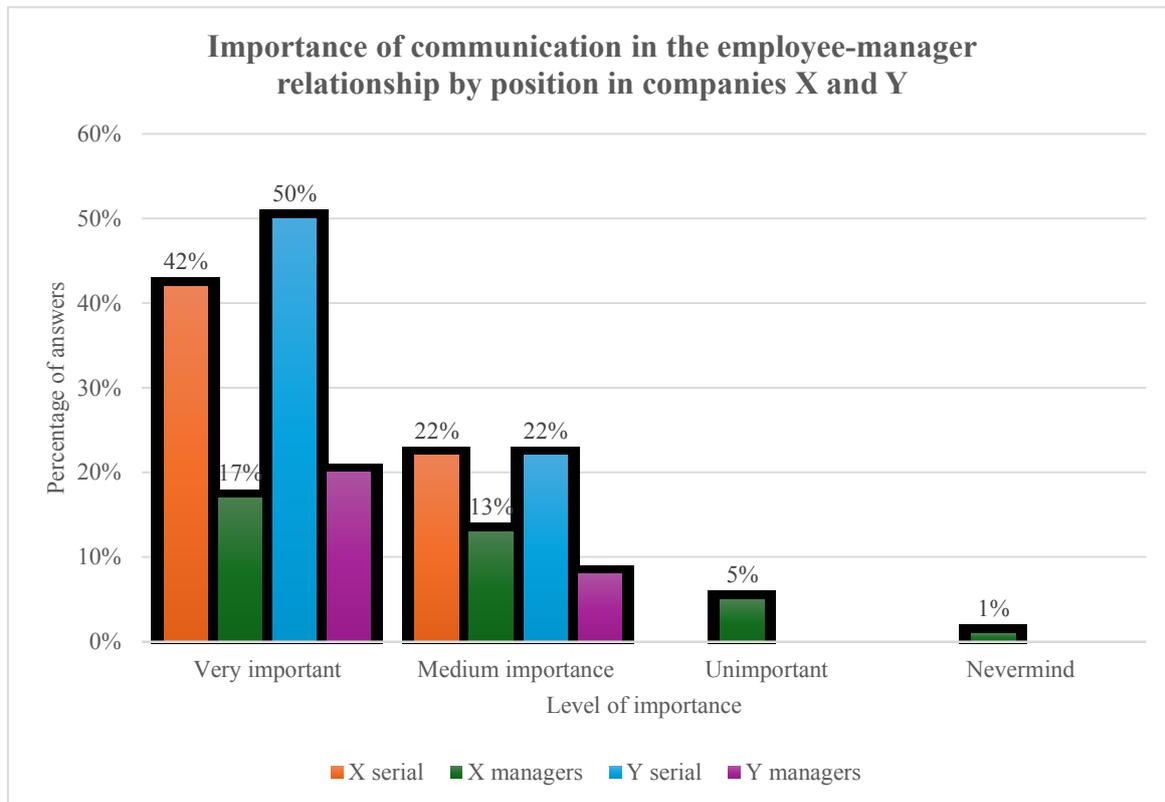

**Figure 8.** Responses of employees of companies that were analysed on the internal communication system to question 8 of the research survey.

Among serial employees, 42% in Company X and 50% in Company Y rated it as "very important," while 17% of managers in Company X and 20% in Company Y selected the same option. Medium importance was chosen by 22% (X serial), 13% (X managers), 22% (Y serial), and 9% (Y managers). The remaining responses show that 5% of managers in Company X marked the relationship as "unimportant," and 1% selected "never mind".

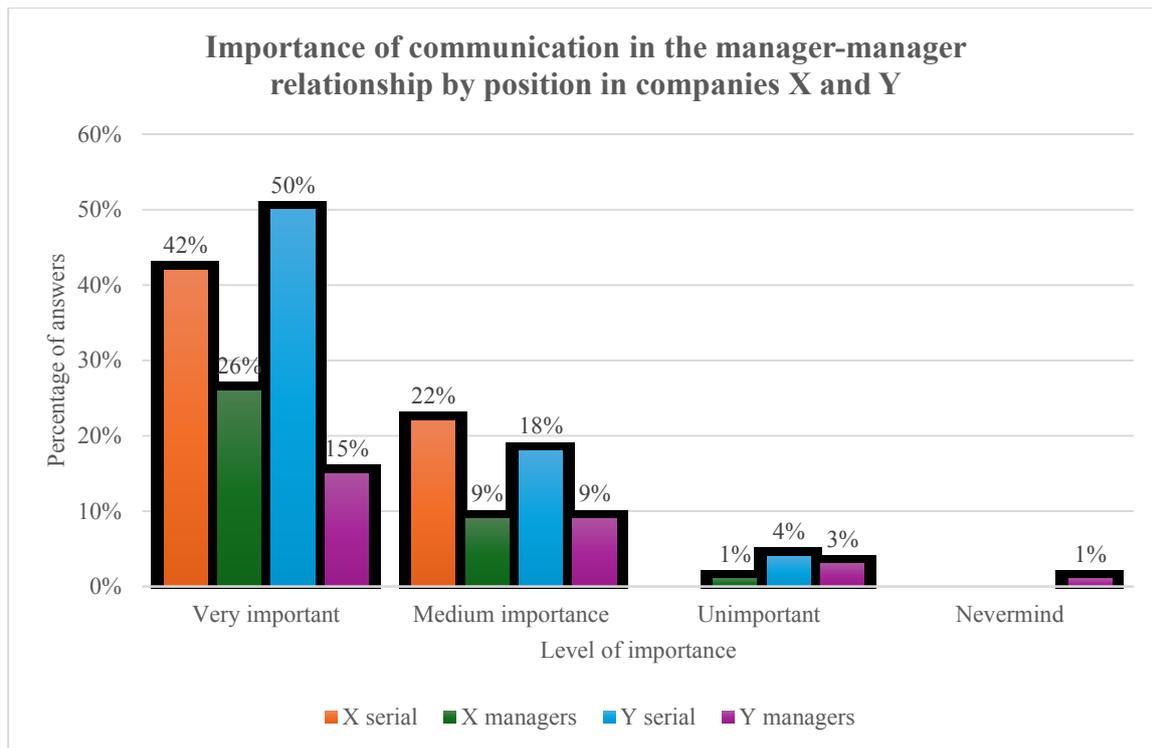

**Figure 9.** Responses of employees of companies that were analysed on the internal communication system to question 9 of the research survey.

Among serial employees, 42 % in Company X and 50 % in Company Y rated it as "very important," while 17 % of managers in Company X and 20 % in Company Y selected the same option. Medium importance was chosen by 22 % (X serial), 13 % (X managers), 22 % (Y serial), and 9 % (Y managers). The remaining responses indicate that 5 % of managers in Company X considered the relationship "unimportant," and 1 % marked "never mind".

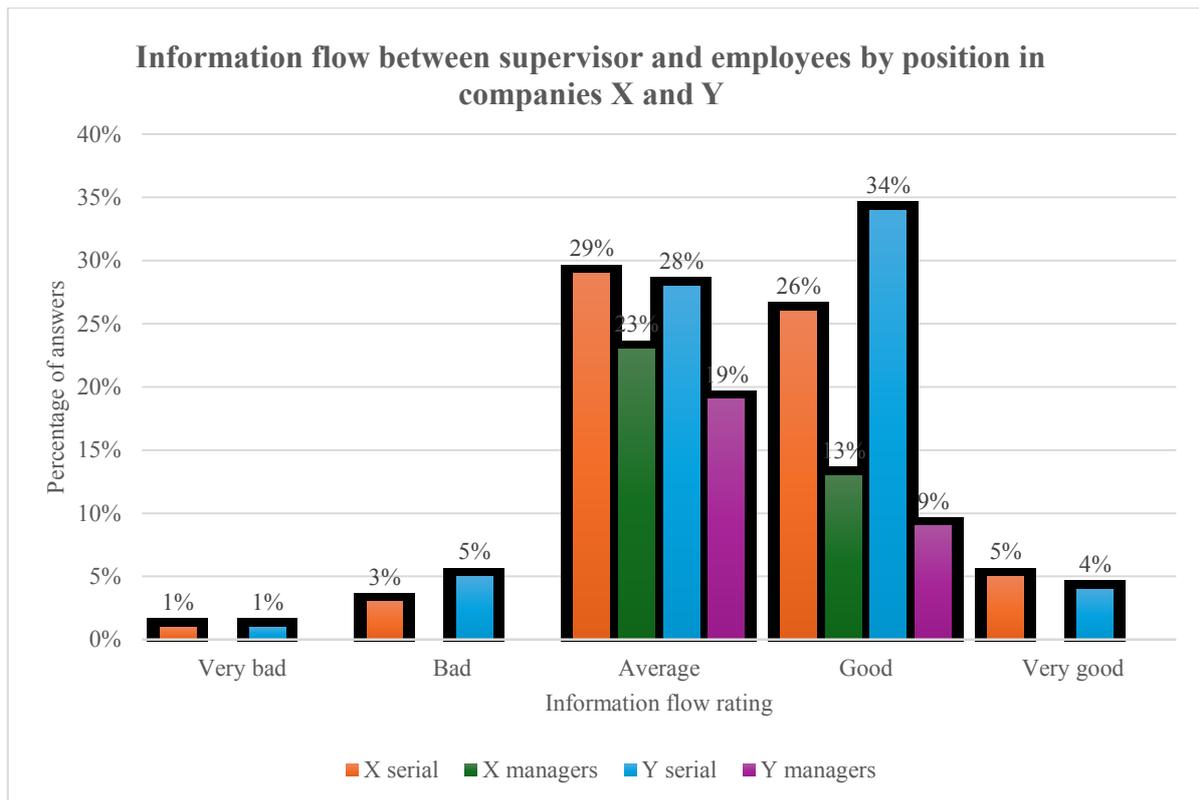

**Figure 10.** Responses of employees of companies that were analysed on the internal communication system to question 10 of the research survey.

Among serial employees, 29% from Company X and 28% from Company Y rated the flow as "average", while 26% (X) and 34% (Y) rated it as "good". For managerial staff, 23% (X) and 19% (Y) considered it "average", and 13% (X) and 9% (Y) as "good". Only a small percentage assessed the information flow as "very bad" or "bad," with the highest negative score (5%) appearing among Y serial employees in the "bad" category. The highest "very good" rating came from X serial employees (5%), followed by Y serial employees (4%). Managers across both companies provided noticeably fewer extreme evaluations, particularly in the "very good" and "very bad" categories.

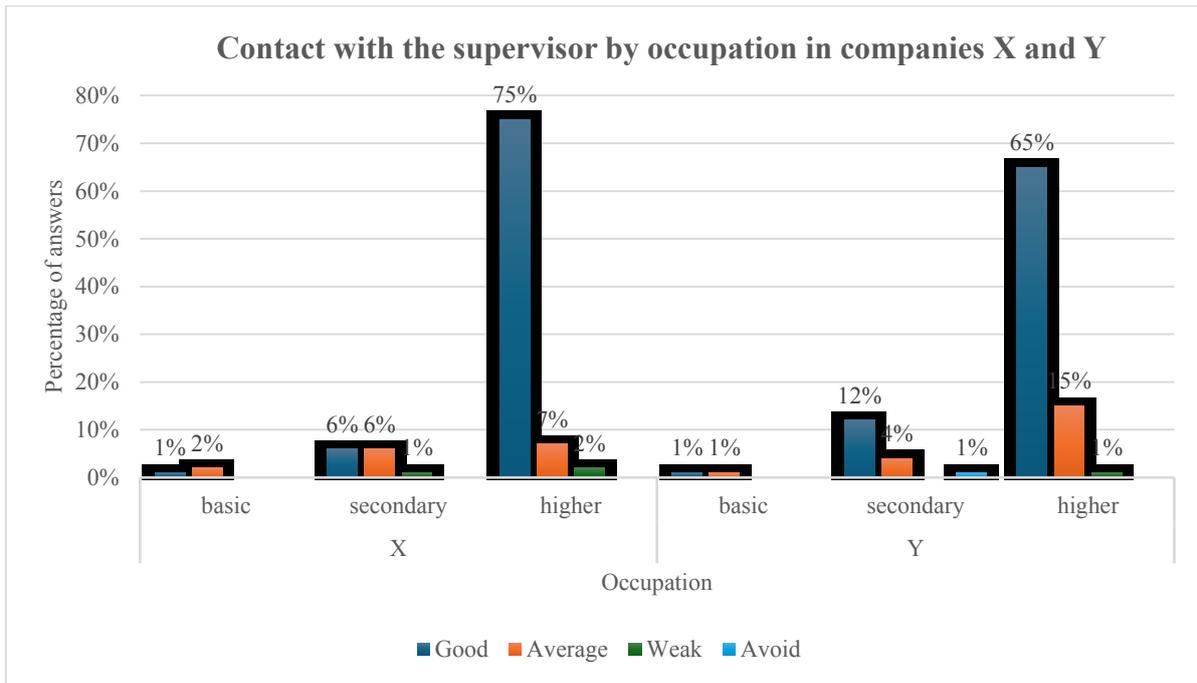

**Figure 11.** Responses of employees of companies that were analysed on the internal communication system to question 11 of the research survey.

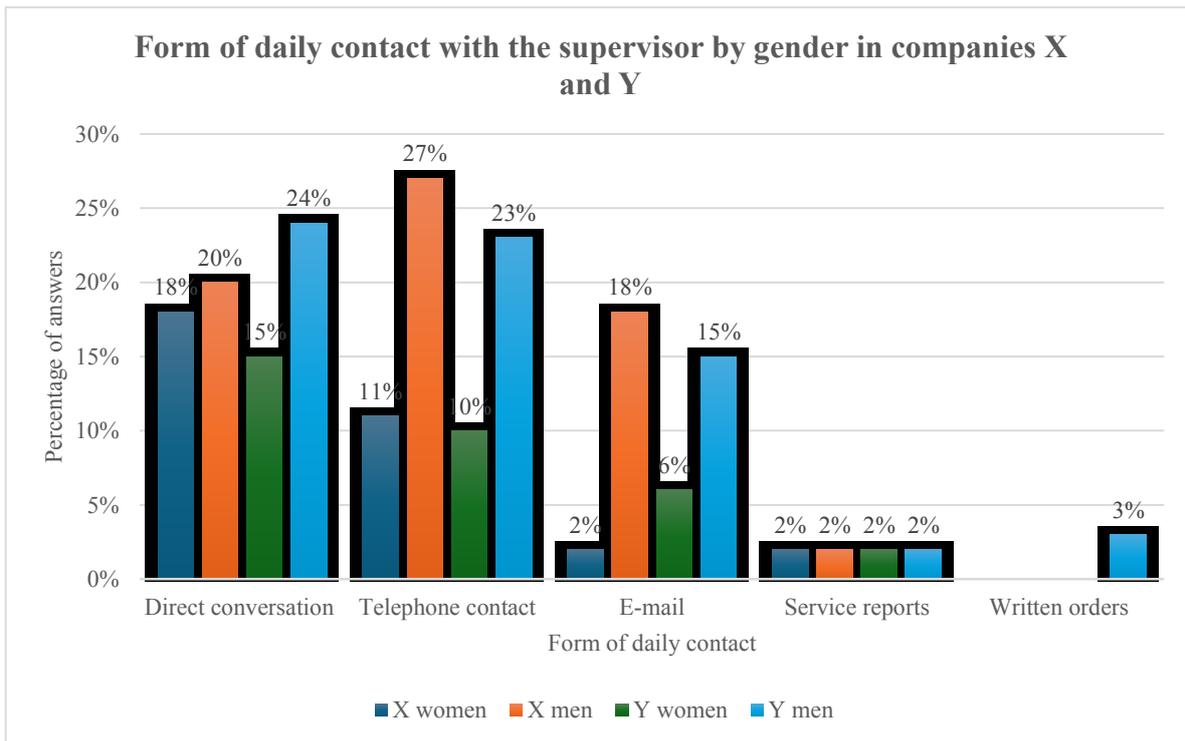

**Figure 12.** Responses of employees of companies that were analysed on the internal communication system to question 12 of the research survey.

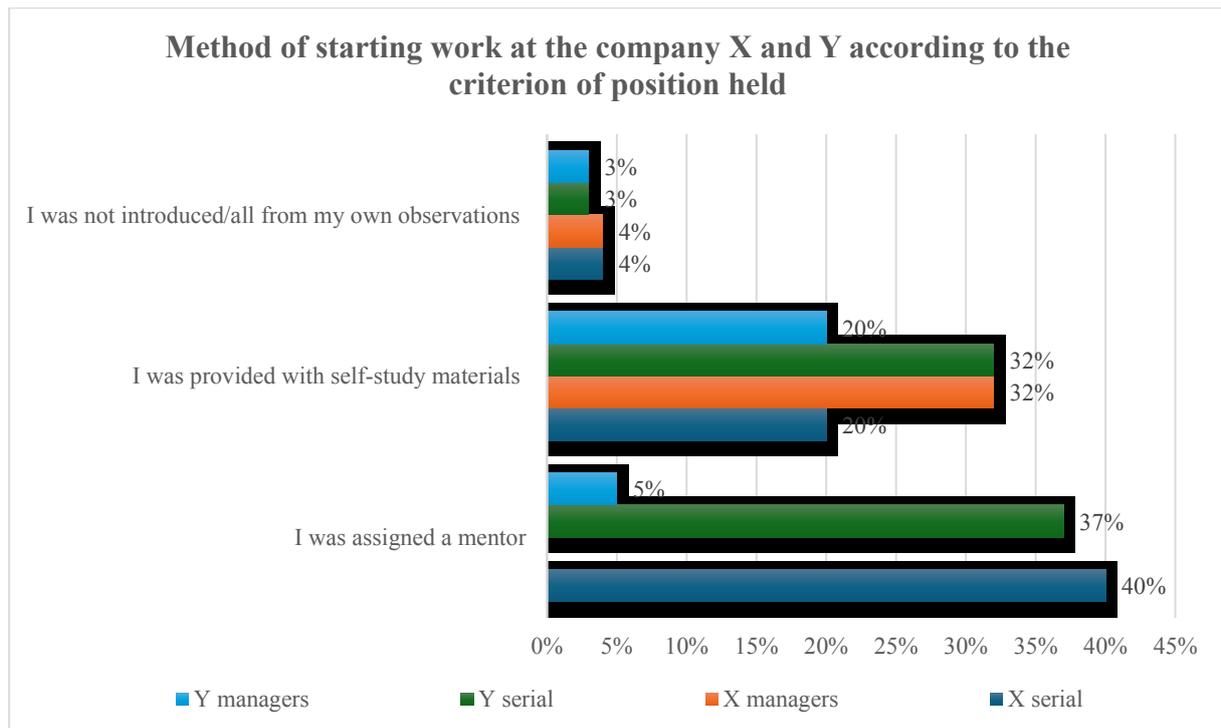

**Figure 13.** Responses of employees of companies that were analysed on the internal communication system to question 13 of the research survey.

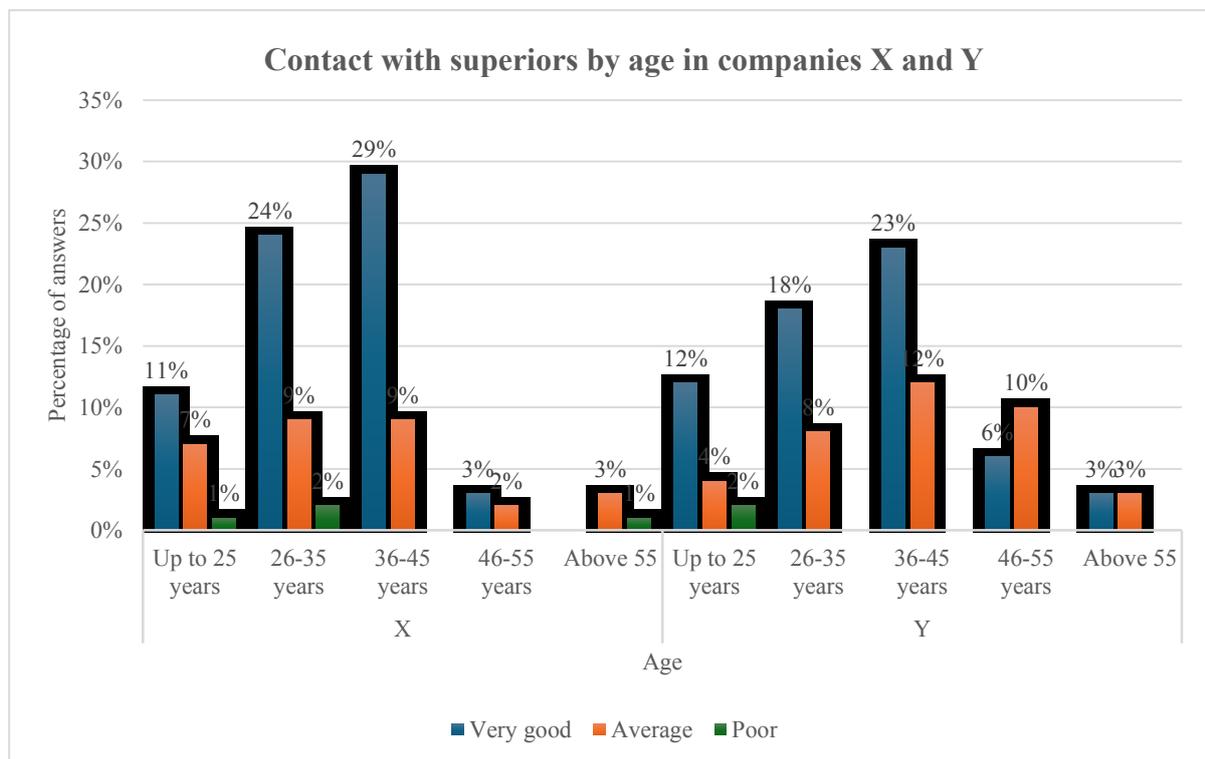

**Figure 14.** Responses of employees of companies that were analysed on the internal communication system to question 14 of the research survey.

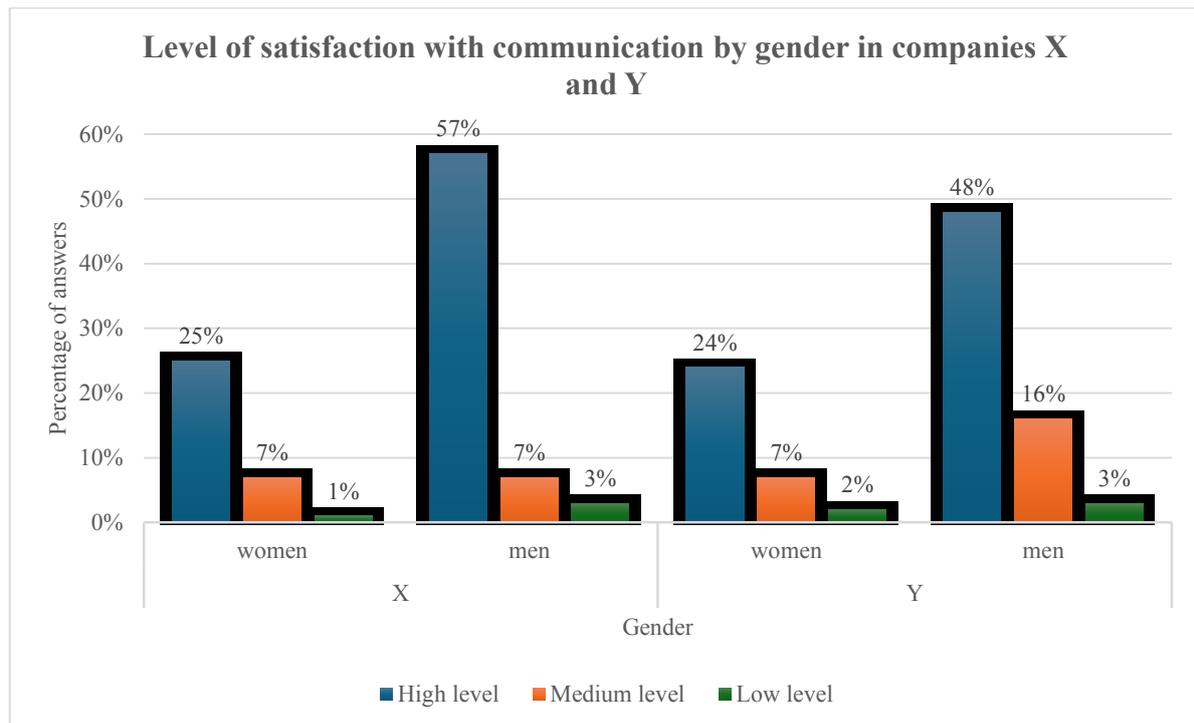

**Figure 15.** Responses of employees of companies that were analysed on the internal communication system to question 15 of the research survey.

Figure 11 illustrates the assessment of contact with the supervisor by occupation level in Companies X and Y. In Company X, 75% of respondents with higher education rated their contact as "good", compared to 65% in Company Y. Respondents with secondary education in Y more often evaluated the contact as "good" (12%) than in X (6%). Very few responses indicated "weak" or "avoid" contact, ranging from 1% to 2% across all education levels and companies.

Figure 12 presents the forms of daily contact with supervisors by gender in both companies. Telephone contact was the most frequent method among X men (27%) and Y men (23%). Direct conversation was used by 20% of X men and 24% of Y men, while X women (18%) and Y women (15%) also indicated similar preferences. Other forms such as service reports and written orders were rarely used, with no value exceeding 3%.

Figure 13 shows how employees started their work in Companies X and Y, based on their position. The most common method for X serial employees (40%) and Y serial employees (37%) was being assigned a mentor. In contrast, 32% of X managers and 32% of Y managers indicated that they were provided with self-study materials. Only a small share of all groups reported no introduction, with values between 3% and 4%.

Figure 14 depicts how employees assessed contact with supervisors by age in Company X. The best ratings ("very good") were given by employees aged 36-45 (29%) and 26-35 (24%). Younger respondents under 25 gave lower ratings, with 11% selecting "very good" and 7% "average". The poorest scores were reported by employees aged 46-55 and above 55, with no category exceeding 3%.

Figure 15 presents the same evaluation of supervisor contact by age group, this time in Company Y. The highest "very good" scores came from employees aged 36-45 (23%) and 26-35 (18%). The lowest scores were observed in the oldest group (above 55), where only 3% indicated "very good". Younger groups (under 25) reported moderate satisfaction, with 12% rating contact as "very good" and 4% as "average".

The analysis of survey results, based on 15 internal communication-related questions, reveals meaningful differences in how employees from Company X and Company Y perceive communication practices within their organizations. The survey covered key aspects such as message clarity, managerial availability, interdepartmental communication, the use of communication tools, and access to feedback.

1. General Perceptions: Employees of Company X consistently rated internal communication more favorably than those of Company Y. Across nearly all questions, Company X outperformed Company Y, particularly in areas such as the clarity of communication, the openness of leadership, and the frequency of updates regarding organizational changes.
2. Communication Tools and Transparency: Company X was perceived as more technologically advanced in its use of internal communication tools (e.g., intranet platforms, internal messaging apps), which enhanced the transparency of managerial decisions. Employees from Company X reported a higher level of understanding of strategic goals and access to up-to-date information.
3. Managerial Accessibility and Feedback Culture: Company X also scored significantly higher in terms of managerial availability and responsiveness. Employees noted that they could easily approach their supervisors and that feedback—both positive and corrective—was provided more regularly and constructively. This reflects a culture of openness and continuous dialogue.
4. Interdepartmental Communication: Survey results highlighted better coordination and information flow between departments in Company X. In contrast, Company Y was characterized by communication silos and limited interaction between units, which affected responsiveness and flexibility in daily operations.
5. Communication Barriers: A notable outlier in the data was Question 10, which assessed the presence of communication barriers. Here, lower scores were favorable, and again, Company X showed better performance, indicating fewer obstacles in the flow of information.
6. Employee Involvement and Organizational Understanding: Company X employees more often reported a sense of inclusion in communication processes and a stronger understanding of organizational changes and expectations. This suggests a more participatory and integrative communication model.

These findings reflect clear differences in the structure and effectiveness of internal communication systems in the two companies. Company X appears to have a more open, transparent, and technologically supported communication environment, contributing to higher employee satisfaction and organizational cohesion.

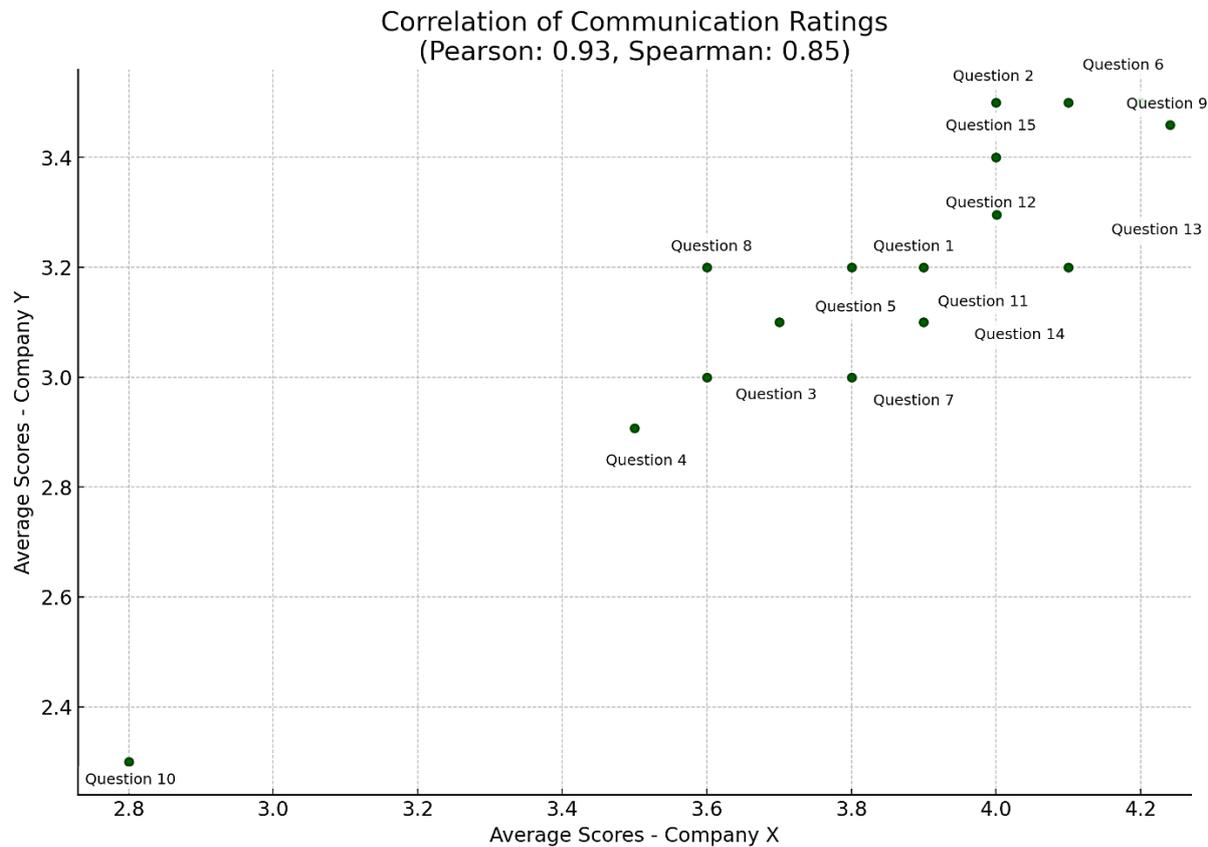

**Figure 16.** Statistical Correlation Analysis Between Company X and Company Y.

The Correlation of Communication Ratings illustrates the relationship between employee perceptions of internal communication practices in Company X and Company Y, based on 15 key survey questions. Each point on the chart represents a communication aspect (e.g., clarity of messages, feedback frequency, managerial availability) and reflects the average rating provided by respondents from both companies. One data point in the chart — Question 10 — is positioned noticeably lower than the others. This is because it relates to communication barriers, a negatively framed item. Unlike the other questions, where higher scores reflect better communication, this one measures the presence of obstacles. Therefore, a lower score indicates a more favorable situation (fewer barriers), which explains its distinct placement on the plot.

Two major statistical coefficients were calculated to assess the strength and nature of the relationship:
- Pearson's correlation coefficient (r = 0.93) indicates a very strong positive linear relationship between the average scores in both companies. This suggests that communication aspects highly rated in Company X tend to be also rated higher in Company Y, although the absolute scores are consistently lower for Company Y.
- Spearman's rank correlation coefficient (ρ = 0.85) confirms a strong monotonic relationship, indicating a consistent ranking of communication aspects between the two organizations, regardless of the exact rating values.

Both correlations are statistically significant with p-values < 0.001, suggesting the observed relationships are unlikely to be due to chance. In addition, the coefficient of variation was used to evaluate the relative variability of responses:
- Company X: 9.06%.
- Company Y: 9.63%.

These values indicate that the dispersion of ratings around the mean is low and similar for both companies. However, Company Y exhibits slightly greater inconsistency in how employees evaluate internal communication practices.

The high degree of correlation, coupled with consistent differences in average scores, implies that although both companies share a similar structure of communication challenges and strengths, Company X systematically outperforms Company Y across all surveyed dimensions. This performance gap is particularly visible in areas such as feedback regularity, message clarity, access to organizational updates, and technological tools used for communication.

## 5. Conclusions and future research implications

The comparative analysis of internal communication practices in Companies X and Y reveals a clear disparity in the quality and effectiveness of their communication systems. Company X demonstrates a significantly higher level of communication effectiveness across all measured dimensions, including clarity of messages, managerial accessibility, use of digital tools, feedback culture, and interdepartmental information flow. Employees in Company X report greater engagement, a stronger sense of inclusion in communication processes, and fewer barriers to information exchange. In contrast, Company Y shows more variability in employee perceptions and exhibits signs of communication silos and a less structured feedback system.

These findings confirm that effective communication is not only a technical process but a cultural and strategic element of organizational life that can influence employee satisfaction, motivation, and overall operational efficiency. Our research establishes new evidence that participatory and technologically supported communication models are associated with higher employee engagement and more positive perceptions of communication efficiency.

The empirical results align with earlier findings by Santos et al. (2024), who emphasize that internal communication significantly affects job satisfaction and motivation. The present study corroborates their model by showing that organizations with more open, regular, and inclusive communication practices (as seen in Company X) tend to foster higher levels of employee engagement and perceived clarity.

Similarly, the outcomes reinforce the CIMO-based study by Eskelinen et al. (2017), where participative communication processes led to improved alignment and motivation. In our study, Company X's use of technology and participatory communication confirms the value of involving employees in the communication ecosystem. However, while Eskelinen's study focused on internal process design within one organizational setting, our comparative approach adds an inter-organizational dimension, exposing how similar industries may adopt divergent communication strategies with varying levels of effectiveness.

Based on the research, several practical implications can be proposed:
- Investment in communication technologies, such as platforms like intranet systems, chat tools, and digital dashboards can enhance message consistency. Company Y could benefit from implementing such tools.
- Implementing a feedback-oriented culture by regular and constructive feedback from supervisors improves trust and employee development.
- Cross-functional meetings, joint workshops, or shared digital spaces could help company Y address its communication silos.
- Implementing participatory communication models, involving employees in designing communication tools or protocols, as shown in the literature and reflected in Company X's practices, increases transparency and motivation.

Despite its contributions, this study is subject to several limitations. Firstly, the scope of the sample is limited, as the research focused exclusively on two companies operating within the same industry and national context. This narrow focus restricts the generalizability of the findings and may not reflect communication dynamics in organizations of different sizes, sectors, or cultural settings.

Secondly, the study relied on self-reported survey data, which inherently carries the risk of response bias. Participants may have answered in a socially desirable manner or misunderstood the intent of certain questions, thereby affecting the accuracy and objectivity of the results.

Finally, the research offers a static, cross-sectional view of internal communication practices. Since data was collected at a single point in time, the study does not capture changes or developments in communication strategies. As a result, it is difficult to assess the sustainability of observed practices or determine how communication effectiveness may evolve in response to internal or external changes.

Future studies could expand and refine the present analysis by exploring internal communication systems across a wider range of industries. Cross-sector comparisons would allow researchers to identify industry-specific challenges and best practices, offering a broader perspective on communication effectiveness. Additionally, conducting longitudinal studies could provide insights into how communication strategies evolve over time, especially in response to organizational changes, leadership transitions, or external disruptions.

Another valuable direction would be to investigate the relationship between internal communication and specific employee outcomes, such as staff turnover, job performance, customer satisfaction, or levels of innovation. Given the increasing popularity of hybrid and remote work models, future research should also examine how digital communication tools influence organizational culture, employee engagement, and well-being.